\documentclass[dvips]{acta}
\usepackage{supertabular,lscape,epsfig}
\usepackage{amssymb}
\usepackage{amsmath}
\usepackage{url}

\usepackage{color}
\usepackage[caption=false]{subfig}
\usepackage{verbatim}

\newcommand{\name}{ASASSN-13co}
\newcommand{\galname}{PGC 067159}
\newcommand{\swift}{{\it Swift}}
\newcommand{\msun}{\ensuremath{\rm{M}_\odot}}
\newcommand{\lsun}{\ensuremath{\rm{L}_\odot}}

\newcommand{\farcs}{\mbox{$.\!\!^{\prime\prime}$}}

\newcommand{\ion}[2]{#1$\;${\small\rmfamily\MakeUppercase{\romannumeral #2}}\relax}
\newcommand{\edit}[1]{\textcolor{black}{#1}}

\SetPages{0}{0}

\SetVol{58}{2016}

\begin{document}

\begin{Titlepage}
\Title{Discovery and Observations of the Unusually Luminous Type-Defying II-P/II-L Supernova ASASSN-13co} 

\Author{T.~W.-S.~~H~o~l~o~i~e~n$^{1,2,3}$,
J.~L.~~P~r~i~e~t~o$^{4,5}$,~~
O.~~P~e~j~c~h~a$^{6,7,8}$,\\
K.~Z.~~S~t~a~n~e~k$^{1,2}$,~~
C.~S.~~K~o~c~h~a~n~e~k$^{1,2}$,~~
B.~J.~~S~h~a~p~p~e~e$^{7,9,10}$,\\
D.~~G~r~u~p~e$^{11,12}$,~~
N.~~M~o~r~r~e~l~l$^{13}$,~~
J.~R.~~T~h~o~r~s~t~e~n~s~e~n$^{14}$,\\
U.~~B~a~s~u$^{1,15}$,~~
J.~F.~~B~e~a~c~o~m$^{1,2,16}$,~~
D.~~B~e~r~s~i~e~r$^{17}$,\\
J.~~B~r~i~m~a~c~o~m~b~e$^{18}$,~~
A.~B.~~D~a~v~i~s$^{16}$,~~
G.~~P~o~j~m~a~n~s~k~i$^{19}$,\\
and~~D.~M.~~S~k~o~w~r~o~n$^{19}$}
{$^1$Department of Astronomy, The Ohio State University, 140 West 18th Avenue, Columbus, OH 43210, USA\\
e-mail: tholoien@astronomy.ohio-state.edu\\
$^2$Center for Cosmology and AstroParticle Physics (CCAPP), The Ohio State University, 191 W. Woodruff Ave., Columbus, OH 43210, USA\\
$^3$US Department of Energy Computational Science Graduate Fellow\\
$^4$N\'ucleo de Astronom\'ia de la Facultad de Ingenier\'ia, Universidad Diego Portales, Av. Ej\'ercito 441, Santiago, Chile\\
$^5$Millennium Institute of Astrophysics, Santiago, Chile\\
$^6$Department of Astrophysical Sciences, Princeton University, 4 Ivy lane, Peyton Hall, Princeton, NJ 08540, USA\\
$^7$Hubble Fellow\\
$^8$Lyman Spitzer Jr. Fellow\\
$^9$Carnegie Observatories, 813 Santa Barbara Street, Pasadena, CA 91101, USA\\
$^{10}$Carnegie-Princeton Fellow\\
$^{11}$Swift MOC, 2582 Gateway Dr., State College, PA 16802, USA\\
$^{12}$Space Science Center, Morehead State University, 235 Martindale Dr., Morehead, KY 40351, USA\\
$^{13}$Las Campanas Observatory, Carnegie Observatories, Casilla 601, La Serena, Chile\\
$^{14}$Department of Physics and Astronomy, 6127 Wilder Laboratory, Dartmouth College, Hanover, NH 03755, USA\\
$^{15}$Grove City High School, 4665 Hoover Road, Grove City, OH 43123, USA\\
$^{16}$Department of Physics, The Ohio State University, 191 W. Woodruff Ave., Columbus, OH 43210, USA\\
$^{17}$Astrophysics Research Institute, Liverpool John Moores University, 146 Brownlow Hill, Liverpool L3 5RF, UK\\
$^{18}$Coral Towers Observatory, Cairns, Queensland 4870, Australia\\
$^{19}$Warsaw University Astronomical Observatory, Al. Ujazdowskie 4, 00-478 Warsaw, Poland}

\Received{Month Day, Year}
\end{Titlepage}

\Abstract{We present photometric and spectroscopic observations of {\name}, an unusually luminous Type II supernova and the first core-collapse supernova discovered by the All-Sky Automated Survey for SuperNovae (ASAS-SN). First detection of the supernova was on UT 2013 August 29 and the data presented span roughly 3.5 months after discovery. We use the recently developed model from Pejcha \& Prieto (2015) to model the multi-band light curves of {\name} and derive the bolometric luminosity curve. We compare {\name} to other Type II supernovae to show that it was unusually luminous for a Type II supernova and that it exhibited an atypical light curve shape that does not cleanly match that of either a standard Type II-L or Type II-P supernova.}{supernovae: general, supernovae: ASASSN-13co}


\section{Introduction}
\label{sec:intro}

Type II supernovae have been widely studied and are known to arise from progenitors that retain their hydrogen envelopes before exploding as core-collapse supernovae (see Filippenko 1997 for a review). Traditionally, these supernovae have been divided into two groups based on the shapes of their light curves: Type II-Linear (II-L), which show a steady decline in magnitude at optical wavelengths, and Type II-Plateau (II-P), which exhibit a lengthy ``plateau'' phase in their optical light curves during which the magnitude decline is very small (e.g., Arcavi et al. 2012; Faran et al. 2014a,b). These observed differences arise from differences in the thickness of the hydrogen envelope remaining around the star (e.g., Sanders et al. 2015). However, in recent years it has been suggested that this distinction is perhaps an oversimplification, and that Type II supernova light curve shapes are simply a continuum of properties (e.g., Anderson et al. 2014; Sanders et al. 2015).

In order to better understand the physics that drive these observed differences, a new phenomenological model has been developed that uses photospheric velocities and multi-band photometry to model the light curves of Type II supernovae and derive physical properties such as bolometric luminosity, ejected nickel mass, and temperature and radius evolution (Pejcha \& Prieto 2015a,b, henceforth jointly referred to as PP15). This model, combined with detailed observations of close-by Type II supernovae, provides a new tool for investigating the diversity of these supernovae and the physical processes that drive their observed properties. In particular, unusual events, such as particularly luminous supernovae or those with light curve shapes that do not match typical Type II-P or Type II-L shapes, may be especially revealing. In addition, the PP15 model provides a method of generating light curves in additional filters beyond those that were used for follow-up observations of Type II supernovae, allowing the comparison of supernovae with disparate datasets. 


\begin{figure}[htb]
\centering
\includegraphics[width=0.98\textwidth]{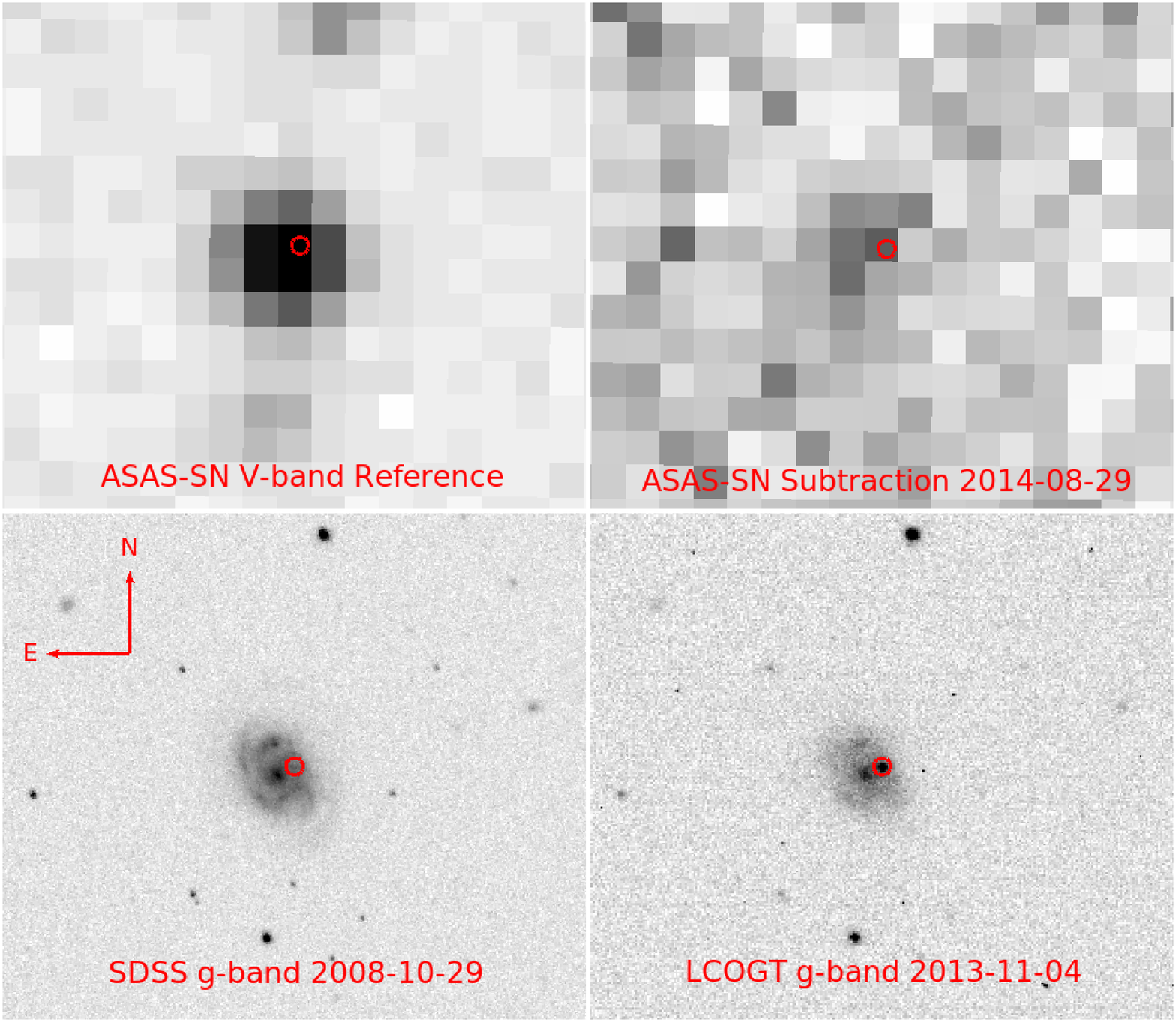}
\FigCap{A finding chart \edit{for} ASASSN-13co. The top-left panel shows the ASAS-SN $V$-band reference image and the top-right panel shows the ASAS-SN subtracted image from 2014 August 29, the date of discovery. The bottom-left panel shows the archival SDSS $g$-band image of the host galaxy {\galname} and the right panel shows an LCOGT $g$-band image taken during the supernova. The dates of the observations are listed on each panel. The red circles have radii of 2\farcs{0} and are centered on the position of the supernova in the LCOGT image.}
\label{fig:finding_chart}
\end{figure}

Here we describe the discovery and follow-up observations of {\name}, an unusually luminous Type II supernova with an atypically slow $V$-band decline rate. The transient was the first core-collapse supernova discovered by the All-Sky Automated Survey for SuperNovae (ASAS-SN\footnote{\url{http://www.astronomy.ohio-state.edu/~assassin/}}; Shappee et al. 2014), a long-term project to monitor the whole sky on a rapid cadence to find nearby supernovae and other bright transients (see Shappee et al. 2014 for details), such as AGN activity (Shappee et al. 2014), extreme stellar flares (Schmidt et al. 2014), outbursts in young stellar objects (Holoien et al. 2014b), and tidal disruption events (Holoien et al. 2014a; Holoien et al. 2016a,b).

Our transient source detection pipeline was triggered on 2013 August 29, detecting a new source with $V=16.9\pm0.1$~mag at the coordinates RA = 21:40:38.72, Dec = $+$06:30:36.98 (J2000; Holoien et al. 2013). The object was also marginally detected on 2013 August 27 at roughly the same magnitude, but is not detected ($V>17$~mag) in data obtained on 2013 August 23 and earlier. A search at the object's position in the Sloan Digital Sky Survey Data Release 9 (SDSS DR9; Ahn et al. 2012) catalog revealed the source of the outburst to be the spiral galaxy {\galname} at redshift $z=0.023063$, corresponding to a luminosity distance of $d=91.6$~Mpc ($H_0=73$~km~s$^{-1}$~Mpc$^{-1}$, $\Omega_M=0.27$, $\Omega_{\Lambda}=0.73$), and that the ASAS-SN source position was offset by roughly 3\farcs{0} from the center of the host galaxy. Follow-up images obtained on 2013 August 30 by J. Brimacombe with a 16-inch RCOS telescope at Coral Towers Observatory (Cairns, Australia) and on 2013 September 3 with the {\swift} UltraViolet and Optical Telescope (UVOT; Roming et al. 2005) confirmed the detection of the transient. Figure~1 shows the archival SDSS $g$-band image of the host and a Las Cumbres Observatory Global Telescope Network (LCOGT; Brown et al. 2013) $g$-band image including the supernova.

The archival SDSS spectrum of the host is that of a late-type star forming spiral galaxy. To obtain magnitudes of the host galaxy, we performed aperture photometry on archival SDSS $ugriz$ images including all galaxy flux and retrieved 2MASS $JHK_S$ magnitudes from the 2MASS Extended Source Catalog (Skrutskie et al. 2006). We then used the code for Fitting and Assessment of Synthetic Templates (FAST v1.0; Kriek et al. 2009) to fit stellar population synthesis (SPS) models to the archival magnitudes of the host galaxy. The fit was made assuming \edit{an $R_V=3.1$ extinction law (Cardelli, Clayton, \& Mathis 1988)}, an exponentially declining star-formation history, a Salpeter IMF, and the Bruzual \& Charlot (2003) models. We obtained a good SPS fit (reduced $\chi_{\nu}^{2}=0.2$), with the following parameters: $M_{*}=(1.7_{-0.7}^{+0.3})\times10^{10}$~{\msun}, age$=7.1_{-4.7}^{+2.9}$~Gyr, and SFR$=1.2_{-0.4}^{+0.1}$~{\msun}~yr$^{-1}$.

A transient classification spectrum obtained on 2013 September 1 with the Wide Field Reimaging CCD Camera (WFCCD) mounted on the Las Campanas Observatory (LCO) du Pont 2.5-m telescope showed a blue continuum and Balmer lines with P-Cygni profiles characteristic of a young Type II supernova. The broad H$\alpha$ line showed a minimum P-Cygni absorption velocity of $\sim12,000$~km~s$^{-1}$ and a possible detection of a faster component with a velocity of $\sim19,000$~km~s$^{-1}$ (Morrell \& Prieto 2013). With an absolute $V$-band magnitude of roughly $-18.1$ at detection, this was atypically luminous for a Type II SN, and we decided to start a follow-up campaign in order to fully characterize this interesting transient. 

In \S2 we describe data taken of the supernova during our follow-up campaign. In \S3 we analyze these data and describe the properties of the transient, using the recent model from PP15 to fit the supernova's multi-band light curves and derive its bolometric luminosity. Finally, in \S4 we compare these properties to those of supernovae in literature to examine the nature of the object.


\section{Observations} 
\label{sec:obs} 

In this section we summarize our new photometric and spectroscopic observations of {\name}.


\subsection{Photometric Observations}
\label{sec:phot}


\begin{figure}[htb]
\centering
\includegraphics[width=0.88\linewidth]{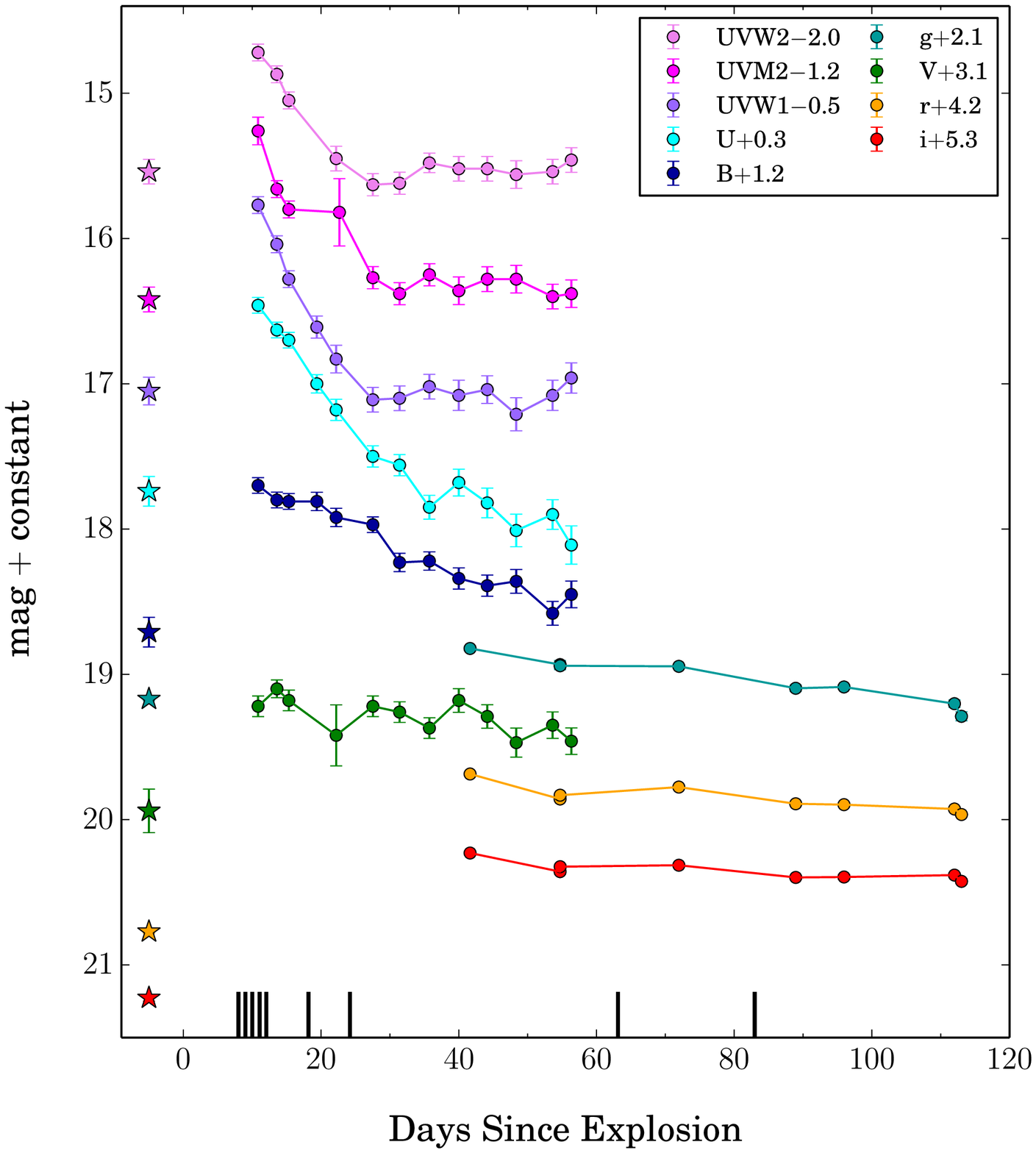} 
\FigCap{Light curves of {\name}, starting at the estimated explosion date (MJD $=56528.1$) and spanning 113 days. Follow-up data obtained from {\swift} (UV $+$ optical) and the LCOGT 1-m (optical) telescopes are shown as circles. All magnitudes are in the Vega system. The data are not corrected for extinction and error bars are shown for all points, but in some cases they are smaller than the data points. Host galaxy magnitudes measured in a 5\farcs{0} aperture centered on the position of the supernova in the subtraction template images are shown as stars at $-5$ days. Dates of spectroscopic follow-up are indicated with vertical black bars at the bottom of the figure. The measurements are heavily contaminated by host galaxy flux in the optical bands.}
\label{fig:lightcurve}
\end{figure}

After detection of the transient, we were granted a series of {\swift} X-ray Telescope (XRT; Burrows et al. 2005) and UVOT target-of-opportunity (ToO) observations between 2013 September 3 and 2013 October 19. We were also granted an additional late-time epoch of observations to be used for host flux subtraction on 2014 April 7. The {\swift} UVOT observations of {\name} were obtained in 6 filters: $V$ (5468~\AA), $B$ (4392~\AA), $U$ (3465~\AA), $UVW1$ (2600~\AA), $UVM2$ (2246~\AA), and $UVW2$ (1928~\AA) (Poole et al. 2008). We used the UVOT software task \textsc{Uvotsource} to extract the source counts from a 5\farcs0 radius region and a sky region with a radius of $\sim$40\farcs0. The UVOT count rates were converted into magnitudes and fluxes based on the most recent UVOT calibration (Poole et al. 2008; Breeveld et al. 2010).

The XRT was operating in Photon Counting mode (Hill et al. 2004) during our observations. The data from all epochs were reduced and combined with the software tasks \textsc{Xrtpipeline} and \textsc{Xselect} to obtain an image in the 0.3$-$10 keV range with a total exposure time of $\sim25,300$~s. We used a region with a radius of 20 pixels (47\farcs{1}) centered on the source position to extract source counts and a source-free region with a radius of 100 pixels (235\farcs{7}) for background counts. From this combined image, we detect a source at the position of the supernova at a level of $5.0_{-1.8}^{+2.1}\times10^{-4}$~counts s$^{-1}$. To convert this to a flux, we assume a power law spectrum with $\Gamma=2$ and Galactic \ion{H}{1} column density (Kalberla et al. 2005), yielding a flux of $\sim 2.5\times10^{-14}$~erg cm$^{-2}$ s$^{-1}$.  At the host distance of $d=92$~Mpc, this corresponds to a luminosity of $L_X=2.5\times10^{40}$~erg s$^{-1}$ ($6.6\times10^6$~{\lsun}). Due to a lack of prior X-ray data for the host galaxy in archival data from Chandra, the X-ray Multi-Mirror Mission (XMM-Newton), or the ROSAT All-Sky Survey (Voges et al. 1999), we cannot determine whether this X-ray flux is from the supernova or the host, and we do not include the X-ray detection in the analyses that follow.

In addition to the {\swift} observations, we obtained $gri$ images with the LCOGT \edit{(Brown et al. 2013)} 1-m telescopes located at Sutherland, South Africa, Cerro Tololo, Chile, and at the MacDonald Observatory between 2013 October 4 and 2013 December 14. We also obtained an image subtraction template epoch in all three filters on 2014 June 5 from the MacDonald Observatory telescope. We measured aperture photometry on the LCOGT images using a 5\farcs0 aperture radius to match the {\swift} UVOT measurements. The photometric zero points were determined using several SDSS stars in the field. All photometric data are shown in Figure~2. 

Figure~2 shows the UV and optical light curves of {\name} from MJD 56528.1, the estimated explosion date \edit{from the PP15 fits} (see \S3), to our latest epoch of observations on MJD 56641.1 (113 days after explosion) without extinction correction or host flux subtraction. Also shown are the dates of spectroscopic follow-up observations and the 5\farcs{0} aperture magnitudes measured at the position of the supernova in the late-time subtraction template images. The $gri$ data \edit{at} all epochs and UV data \edit{at} later epochs are heavily contaminated by host galaxy flux.

In order to isolate the supernova flux for more accurate photometric measurements, we performed host flux subtraction in two ways. For the LCOGT $gri$ images, we aligned the 2014 June 5 template images to the supernova images using stars in the field and used \textsc{Hotpants}\footnote{\url{http://www.astro.washington.edu/users/becker/v2.0/hotpants.html}}, an implementation of the Alard (2000) algorithm for image subtraction, to subtract the templates from the science images. We then performed photometry using a 5\farcs{0} aperture on the subtracted images to obtain host-flux subtracted measurements of the supernova flux. For the {\swift} data, we were unable to use \textsc{Hotpants} because there are too few stars in the field. Instead, we measured the 5\farcs{0} host galaxy flux at the position of the supernova in the 2014 April 7 template image. We then subtracted this flux from the flux measurements made in the science epochs, producing host-subtracted flux measurements that were then converted to host-subtracted magnitudes. These host-subtracted magnitudes were then corrected for Galactic extinction and are used in the light curve fits and analysis presented in \S3.


\subsection{Spectroscopic Observations}
\label{sec:spec}

We obtained nine low- and medium-resolution optical spectra of ASASSN-13co spanning more than two months between 2013 September 1 and 2013 November 15. The spectra were obtained with WFCCD mounted on the LCO du Pont 2.5-m telescope (range $3700-9500$~\AA, $\rm R\sim 7$~\AA), the MDM Modular Spectrograph (Modspec) mounted on the MDM Observatory Hiltner 2.4-m telescope (range $4660-6730$~\AA, $\rm R\sim 4$~\AA), the Ohio State Multi-Object Spectrograph (OSMOS; Martini et al. 2011) mounted on the MDM Observatory Hiltner 2.4-m telescope (range $4200-6800$~\AA, $\rm R\sim 4$~\AA), and with DIS mounted on the Apache Point Observatory 3.5-m telescope (range $3500-9600$~\AA, $\rm R\sim 7$~\AA). The spectra were reduced using standard techniques in IRAF. We applied telluric corrections to all the spectra using the spectrum of the spectrophotometric standard observed the same night. Figure~3 shows a montage of the flux-calibrated spectra.


\begin{figure*}
\begin{minipage}{\textwidth}
\centering
\subfloat{{\includegraphics[width=0.48\textwidth]{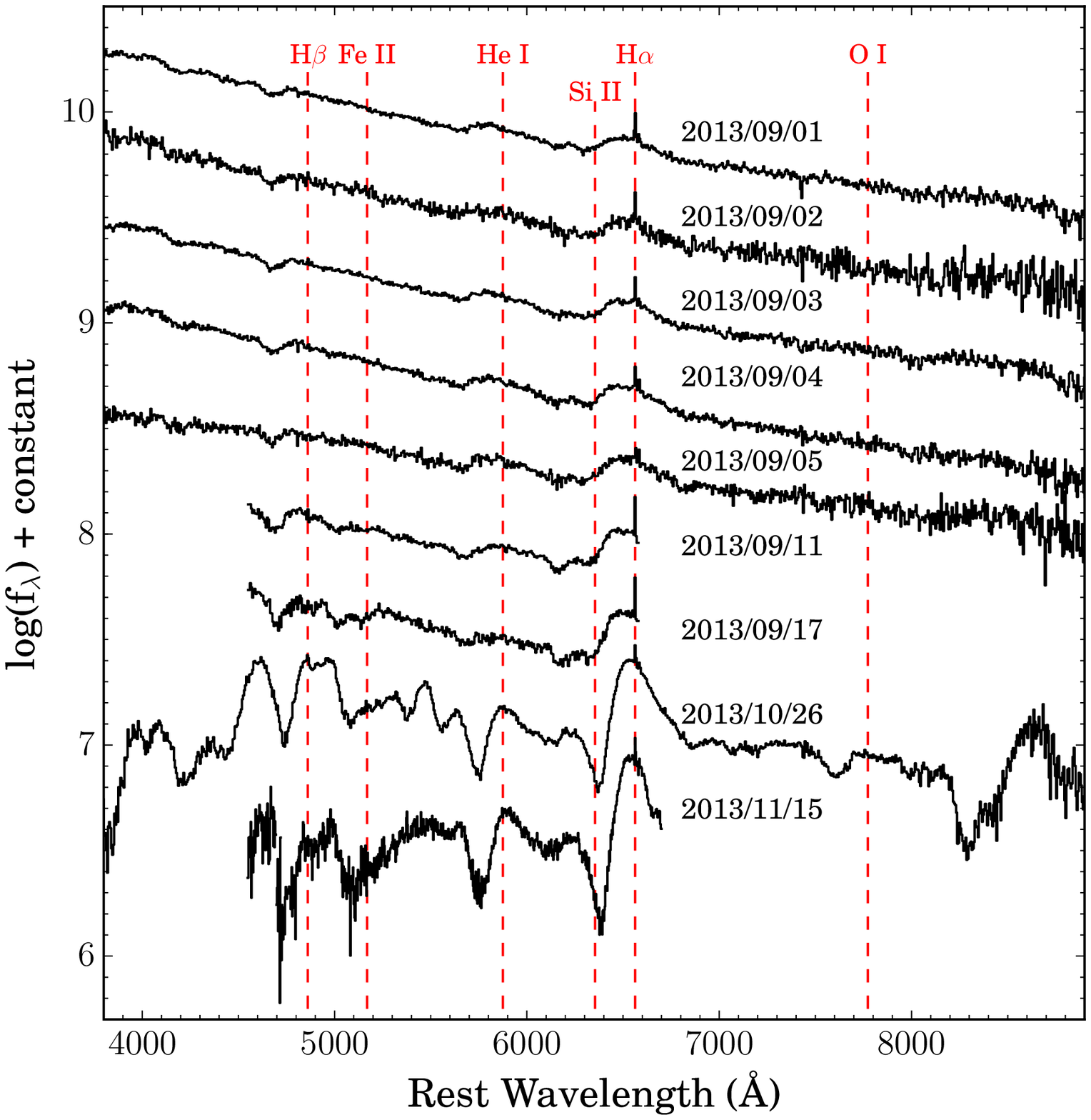}}}
\subfloat{{\includegraphics[width=0.48\textwidth]{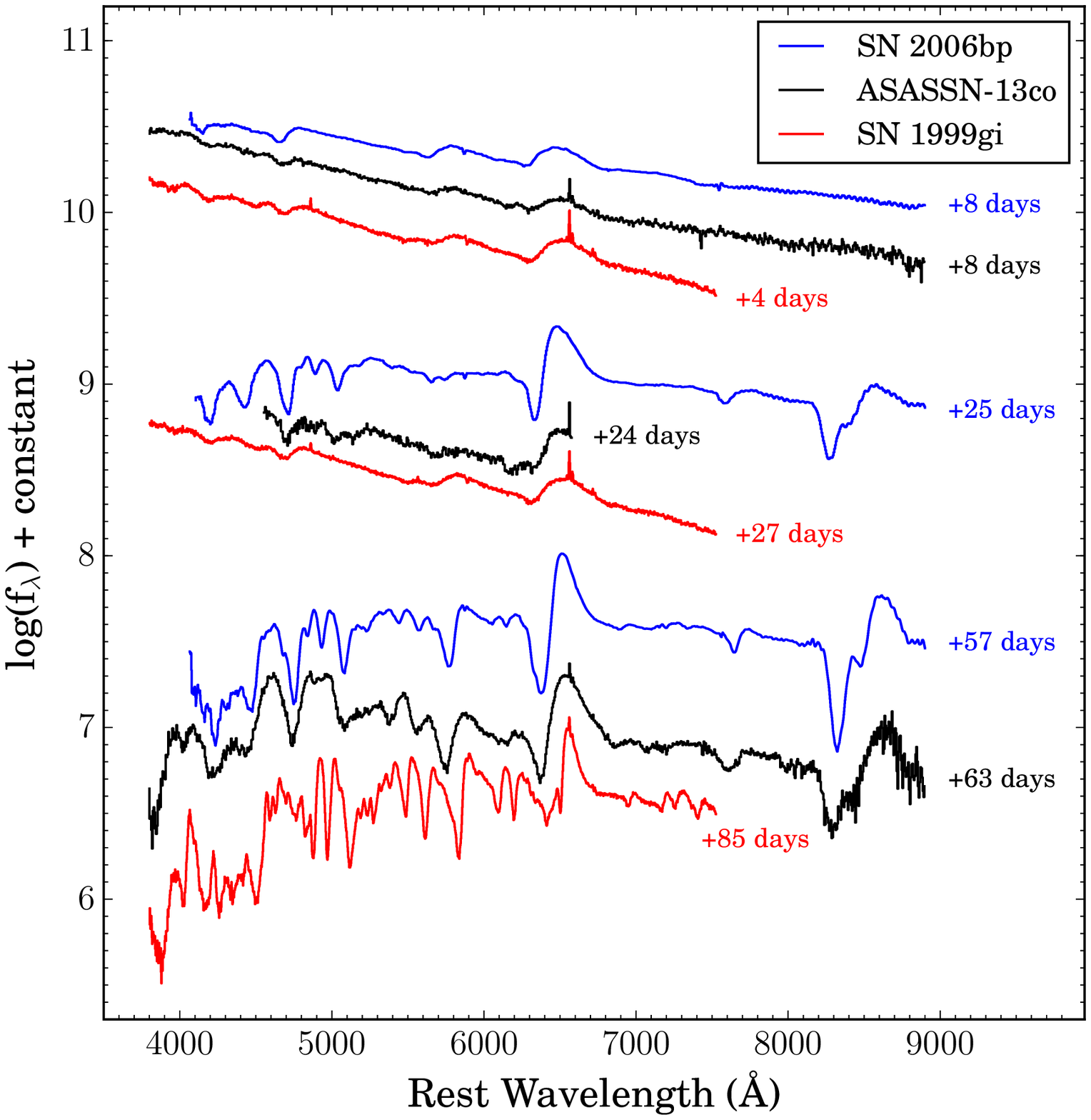}}}
\FigCap{\emph{Left Panel}: Spectral time-sequence for {\name}. Each spectrum \edit{is labelled by the UT date on which} it was obtained. Emission lines used to measure velocities in Figure~4 are labelled and marked with vertical dashed lines. The rapidly fading blue continuum and broad hydrogen emission features are typical of a Type II supernova. \emph{Right Panel}: Spectral comparisons between {\name} and two Type II-P supernovae, SN 2006bp (Quimby et al. 2007) and SN 1999gi (Leonard et al. 2002) at three epochs. The evolution of {\name} looks very similar to both comparison objects, particularly SN 1999gi.}
\label{fig:spectra}
\end{minipage}
\end{figure*}

The main characteristics of the spectra of {\name} are a strong blue continuum in the early epochs that becomes weaker over time and broad emission features in all epochs, most notably H$\alpha$ and H$\beta$, with no narrow emission lines that would indicate circumstellar interaction. These features are characteristic of a Type II supernova, and we used the first follow-up spectrum of {\name} (from UT 2013 September 01) to classify the supernova using the Supernova Identification code (SNID; Blondin \& Tonry 2007). The best SNID matches were normal Type II-Plateau (II-P) supernovae around maximum light and at a redshift $z=0.023$, consistent with the host galaxy redshift measurement in Springob et al. (2005). Based on the SNID results and non-detection limits from ASAS-SN, we constrained the explosion date of {\name} to MJD 56529.1$\pm1$~day. This constraint is used in \S3 when fitting the supernova light curve with the PP15 model.

We compare the spectra of {\name} at three epochs to two Type II supernovae that showed good matches with {\name} in SNID, SN 2006bp (Quimby et al. 2007) and SN 1999gi (Leonard et al. 2002) in Figure~3. All archival spectra of comparison objects were retrieved from the Weizmann Interactive Supernova data REPository (WISEREP; Yaron \& Gal-Yam 2012). The comparison confirms that, spectroscopically, {\name} resembles a normal Type II-P supernova.


\begin{figure}[htb]
\centering
\includegraphics[width=0.95\linewidth]{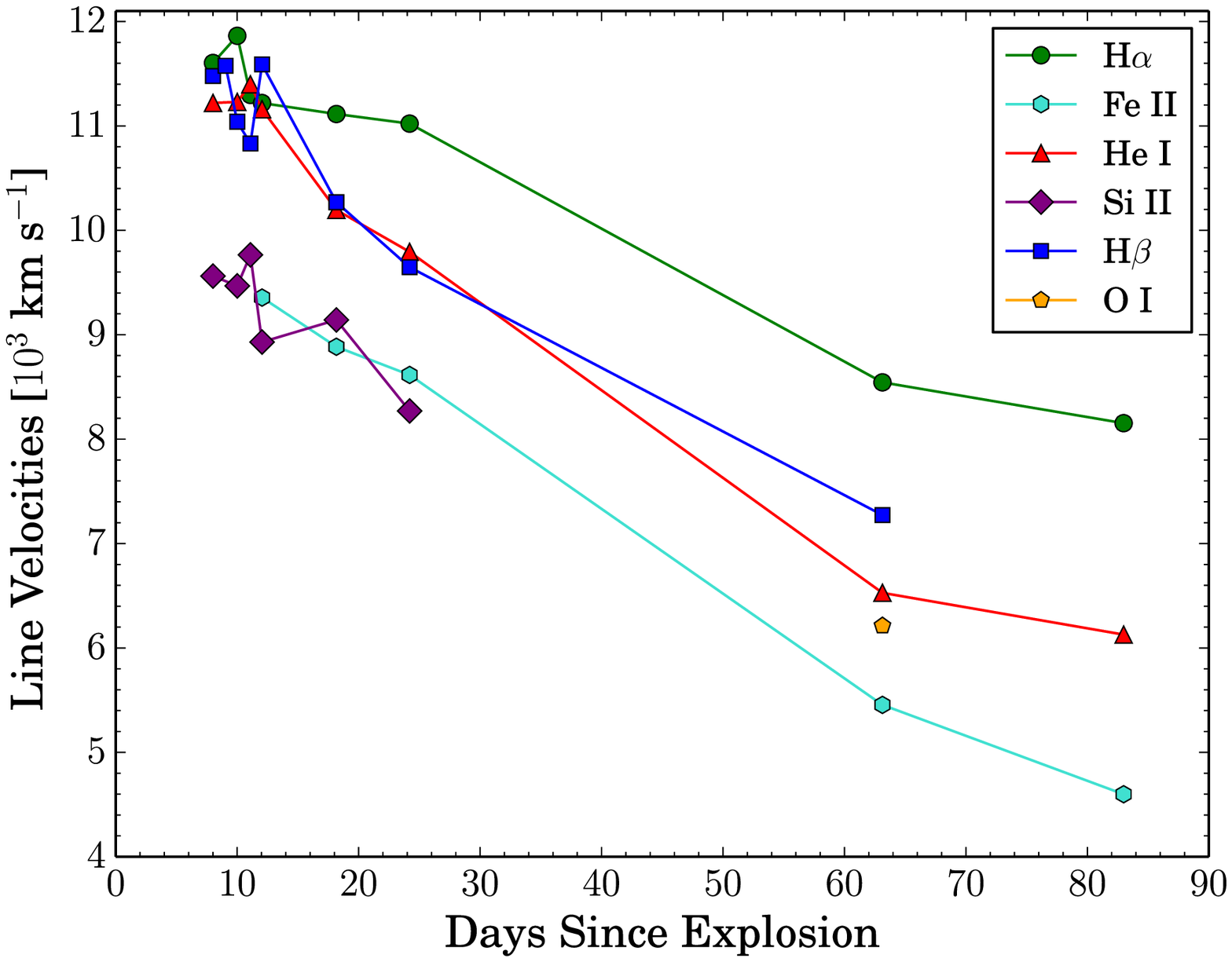}
\FigCap{Velocity evolution for the emission lines used to measure velocities. Each line is shown with a different symbol and color. Symbols represent the measurements, while the connecting lines are provided only to guide the eye. The velocities are higher than typical for a Type II supernova, but the evolution is normal. Table~3 contains all the measured velocity data.}
\label{fig:velocities}
\end{figure}

Finally, we used a number of emission lines in \name's spectra (indicated in the left panel of Figure~3) to measure photospheric velocities at different epochs. Figure~4 shows the velocity measured for each line in each epoch it was detected. The line velocity evolution is consistent with that of other Type II supernovae (e.g., Quimby et al. 2007, Valenti et al. 2014), providing further evidence that, spectroscopically, {\name} does not show any features that differentiate it from normal Type II-P supernovae.


\section{Light Curve Fits and Analysis}
\label{sec:lcanal}

Using the host-subtracted photometry and spectroscopic \ion{Fe}{2} (5169~\AA) line expansion velocities we fit the multi-band light curves of {\name} using the PP15 model. This phenomenological model uses photometric measurements and line velocities to derive the photospheric radius and temperature variations of the supernova. These variations can then be used to calculate light curves in other filters, bolometric luminosities, Nickel-56 masses, and other quantities of interest. The model is based on a sample of 26 well-observed Type II-P supernovae and is designed for fitting such supernovae, but PP15 show that it works for Type II-L supernovae as well. When fitting the light curves of {\name}, we constrained the explosion date to MJD 56529.1$\pm1.0$~day so that the fits will be consistent with our spectroscopic observations. We performed the fit with the distance modulus unconstrained, finding a best-fit value of $34.46\pm 0.23$, which is roughly 0.4 magnitudes off from the value of 34.81 obtained assuming the redshift of 0.023063 from Springob et al. (2005). However, the small number of expansion velocities and the photometric data for {\name} indicate that the distance determined using the expanding photosphere method implemented in PP15 might be unreliable. Therefore, we also fit the light curve with a fixed distance modulus of 34.81. There are no noticeable differences in the light curve fits, and we use these results, \edit{including the best-fit explosion date of MJD 56528.1,} in the analyses that follow. Figure~5shows the host-subtracted photometry and the light curve fits from the PP15 model. The best-fit parameters from the PP15 model are given in Table~1. 


\begin{figure}[htb]
\centering
\includegraphics[width=\textwidth]{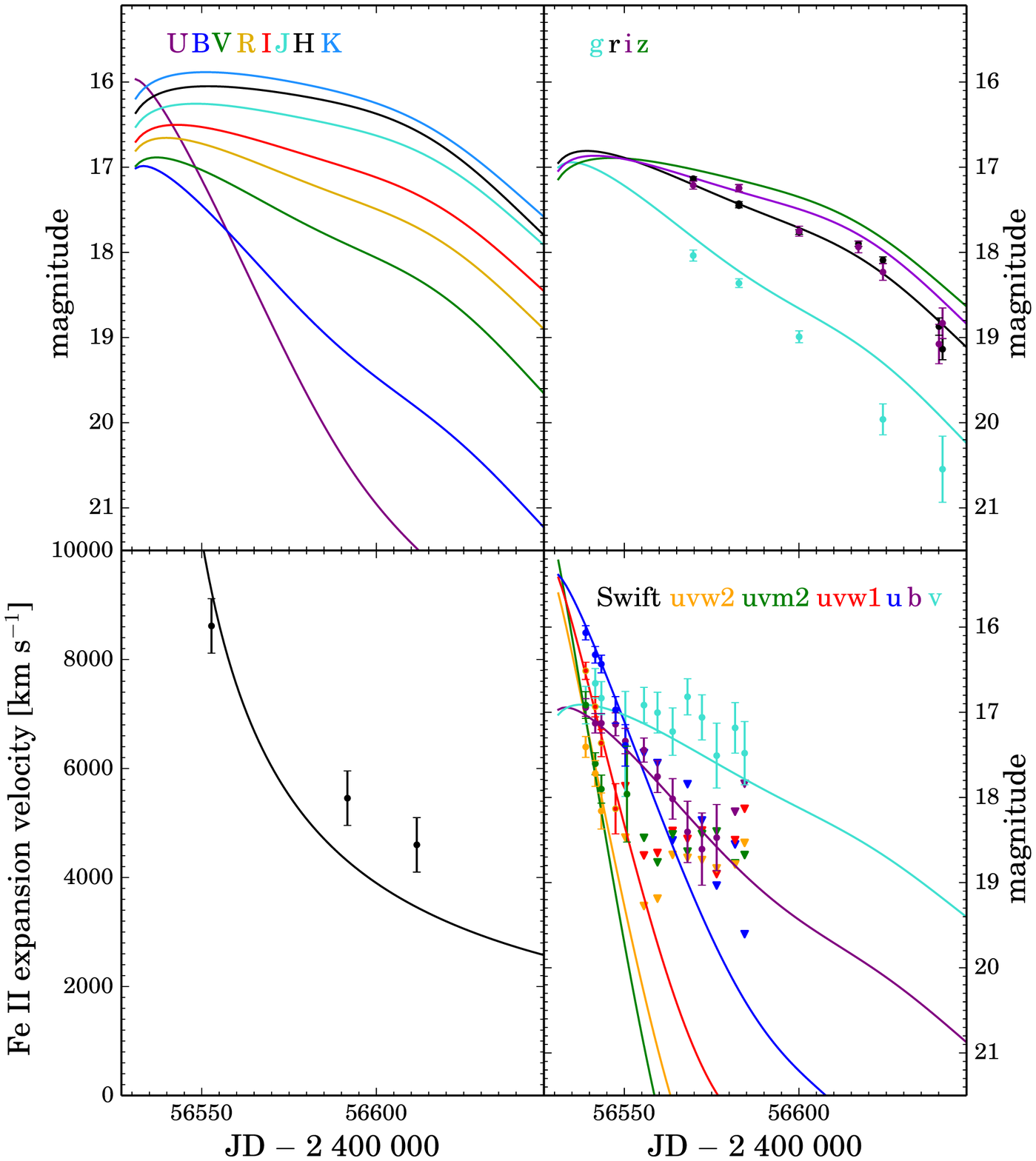}
\FigCap{Host-subtracted light curves and \ion{Fe}{2} velocities of {\name} in various filters from the PP15 model fits (colored lines) and the corresponding host-subtracted photometry and velocity measurements (colored points) covering roughly 3.5 months after the explosion (MJD 56528.1, shortly after the final ASAS-SN non-detection on MJD 56527.9). 3-sigma upper limits are shown as downward-pointing triangles. Error bars are shown for all points, but in some cases they are smaller than the data points. Filters are indicated in each panel with the color of the text matching that of the corresponding light curve. The model seems to fit all bands fairly well despite a lack of wavelength coverage, and we truncate the fits shortly after our last epoch of photometric data, as later epochs are poorly constrained by the model. Based on these fits, {\name} appears to resemble a Type II-L supernova more than a Type II-P, though it declines more slowly than a typical Type II-L. Table~2 contains all the host-subtracted photometric data.}
\label{fig:lcfits}
\end{figure}


\MakeTable{lccc}{12.5cm}{PP15 Model Parameters for {\name}}
{\hline
Parameter & Value & Uncertainty & Rescaled \\
\hline
Explosion time ($t_{exp}$) & 56528.1 & --- & --- \\
Plateau duration ($t_p$) & 84.650 & 3.674 & 5.922 \\
Transition width ($t_w$) & 26.263 & 1.450 & 2.576 \\
$\omega_0$ ($10^3$~km~s$^{-1}$) & 124 & 11 & 14 \\
$\omega_1$ & $-0.809$ & 0.033 & 0.052 \\
$\omega_2$ (km~s$^{-1}$) & 0.0 & --- & --- \\
$\gamma_0$ & $-0.0044$ & 0.0005 & 0.0006 \\
$\gamma_1$ & 11.30 & 0.06 & 0.11 \\
$\alpha_0$ & $-0.0011$ & 0.0002 & 0.0004 \\
$\alpha_1$ & 0.12 & 0.00 & 0.01 \\
Distance modulus & 34.81 & 0.00 & 0.00 \\
E($B-V$) & 0.28 & 0.03 & 0.05 \\
\hline
\multicolumn{4}{p{9cm}}{Variable names correspond to those defined in PP15. The ``Rescaled'' values are the uncertainties scaled so that $\chi^2$/DOF$=$1, as described in PP15. An uncertainty of 0.0 indicates that the parameter's value was fixed. In the cases of $t_{exp}$ and $\omega_2$, the model converged on the lower bound of the allowed range for the parameter, and the uncertainty is not calculated. Given the photometric data constraints, the uncertainty on $t_{exp}$ is about 2 days, but this uncertainty is not propagated into the other parameters.}
\label{table:pp_params}
}

The shape of \name's light curve appears to resemble that of a Type II-L supernova more than a Type II-P. However, the derived model parameters shown in Table~1 do not seem to fall nicely into either category. The II-P supernovae in the PP15 sample typically showed plateau durations of $85$~days~$\lesssim t_p \lesssim 130$~days and transition widths of $1$~day~$\lesssim t_w \lesssim6$~days, while the Type II-L supernova SN 1980K had a plateau duration of $t_p\approx 61$~days and a transition width of $t_w \approx13$~days. {\name} appears to fall somewhere between these two classes, showing both a fairly long plateau phase (though one still shorter than the smallest for the II-P sample) and a long transition width, longer even than that of SN 1980K. We note that the exact values of both the plateau duration and transition width are not well-constrained due to the lack of later photometric data, but the best-fit values are consistent with the light curve shape. This seems to indicate that {\name} is neither a Type II-P nor a Type II-L supernova. As discussed later in this section, this may further illustrate that Type II supernovae exhibit a continuum of morphologies rather than two distinct types. 


\begin{figure}[htb]
\centering
\includegraphics[width=0.95\linewidth]{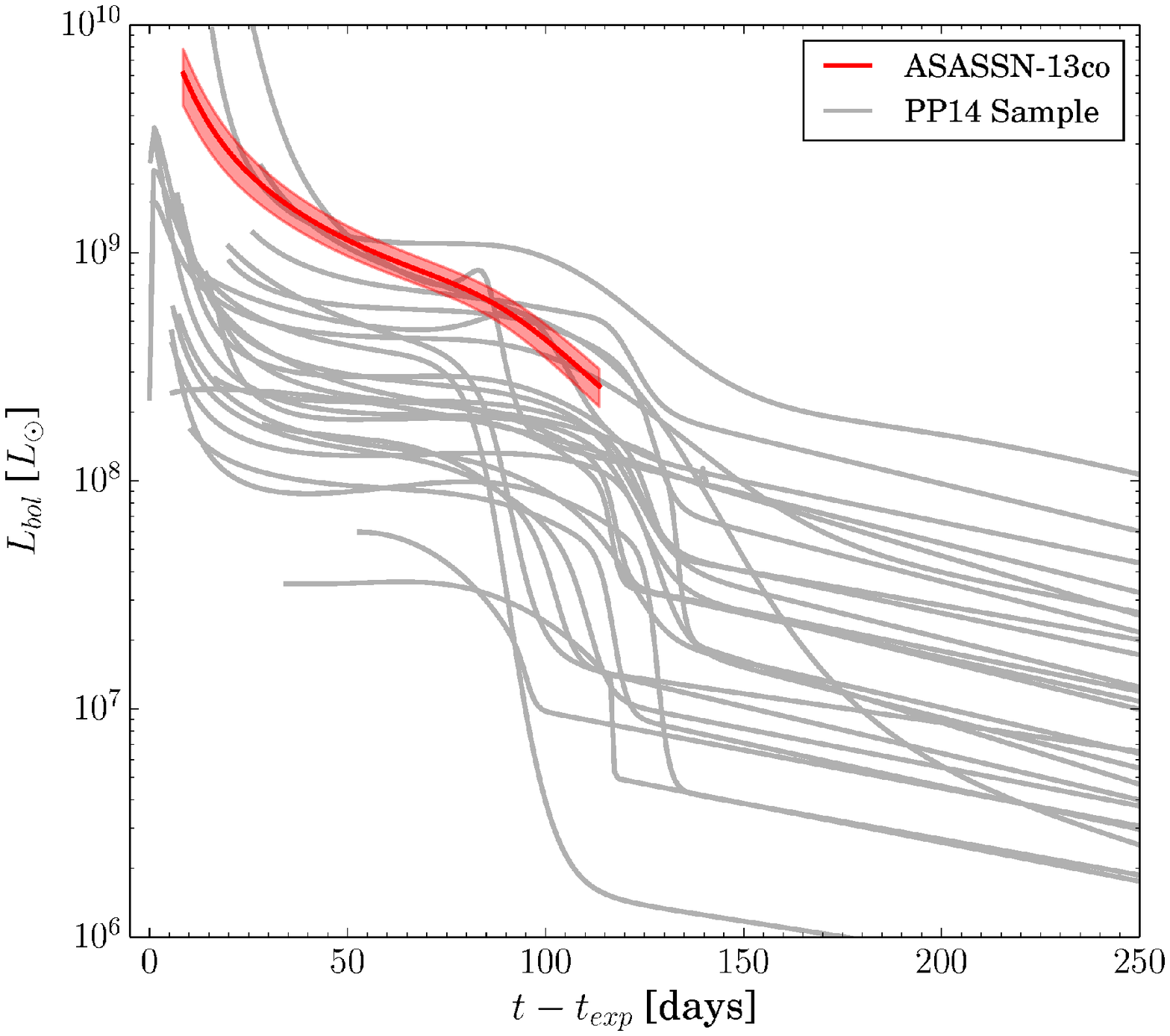}
\FigCap{Evolution of the bolometric luminosity for {\name} (red) compared to the supernovae used in Figure~13 of PP15 (grey). The 1$-\sigma$ uncertainty region is shown in red around the curve. We truncate the luminosity curve for {\name} at the time of our final epoch of photometric data to avoid extrapolation by the model. At early times in particular, {\name} is one of the most luminous supernovae of those shown, and its luminosity decline does not show a strong plateau phase.}
\label{fig:lbol}
\end{figure}

We also obtain an estimate of the evolution of the bolometric luminosity from the PP15 model, which we compare to the supernova sample from PP15 in Figure~6. The PP15 model constructs the bolometric curve of the input supernova by calculating reddening-corrected spectral energy distributions (SEDs) at many epochs from the derived multi-band light curves, and then integrating the flux across all wavelengths in these SEDs using the trapezoidal rule. They cut off the integral at wavelengths shorter than 0.19 $\mu$m and estimate the long wavelength flux by extrapolating the $K$-band flux to infinity assuming the emission follows the Raleigh-Jeans tail of a blackbody. In Figure~6 we truncate the luminosity evolution of {\name} at the date of our last photometric epoch to avoid extrapolation to epochs that are not constrained by data. As can be seen in Figure~6, {\name} is more luminous than all but two of the supernovae in the PP15 sample at early times, and its luminosity decreases steadily, rather than showing the rapid fall and "leveling off" characteristic of the II-P supernovae in the sample. This again indicates that \name's light curve morphology does not resemble that of a Type II-P supernova, despite the fact that its strongest spectral matches from SNID were all to Type II-P supernovae. In addition, Figure~6 shows that {\name} had an atypically high luminosity at maximum, and it could be even higher at earlier epochs, when emission in the UV was strongest and there might have been significant emission at wavelengths shorter than the 0.19 $\mu$m cutoff used in the luminosity integral.


\begin{figure}[htb]
\centering
\includegraphics[width=0.98\linewidth]{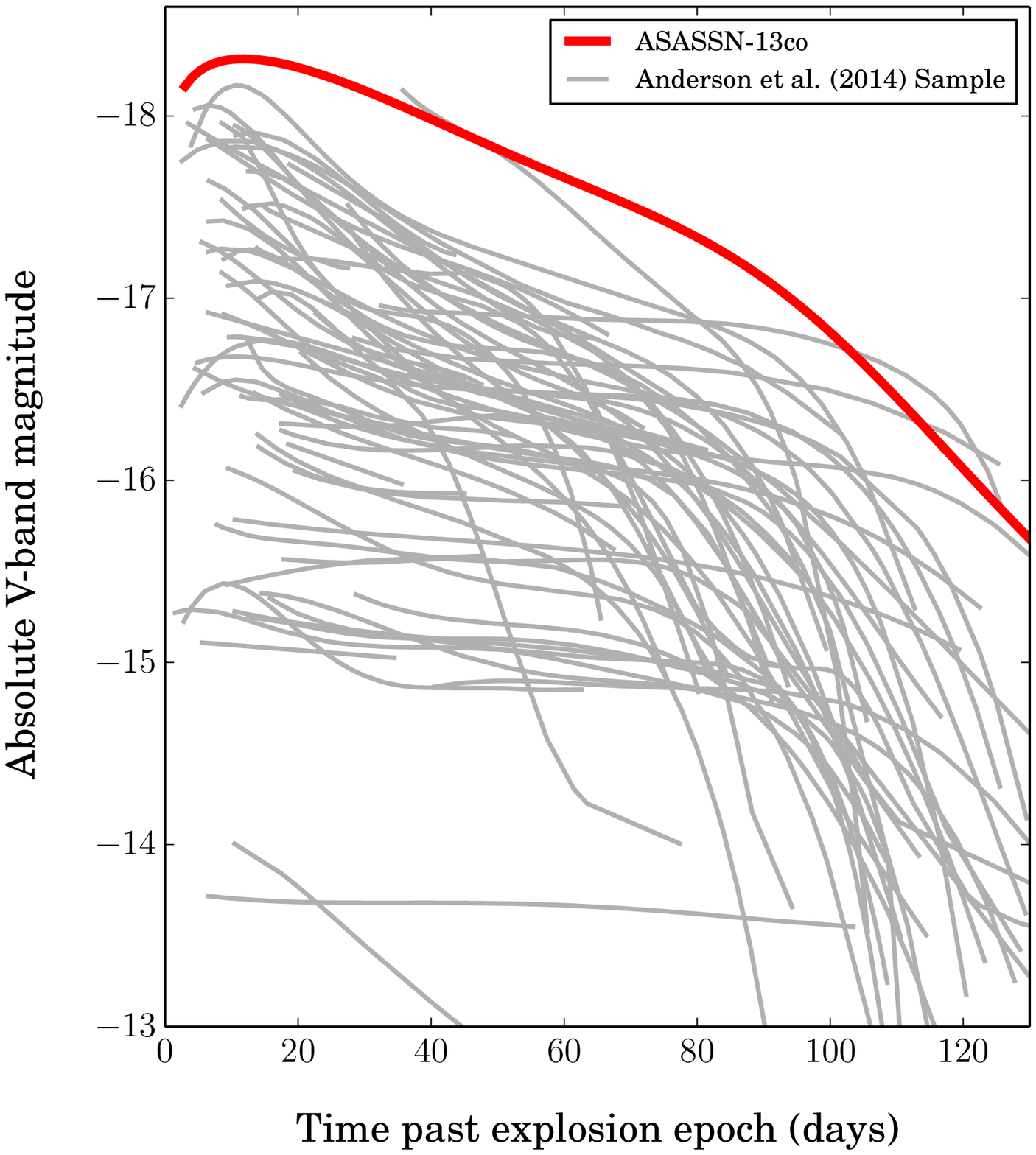}
\FigCap{Comparison of the absolute $V$-band light curve of {\name} (red) to those of the Type II supernova sample shown in Figure~2 of Anderson et al (2014) (grey). The light curve for {\name} is derived from the PP15 model fit and assumes a distance modulus of 34.81. All light curves shown are corrected for Galactic extinction but not for host galaxy extinction. {\name} has one of the most luminous $V$-band light curves of all the supernovae shown, and appears to have a shallower decline than many of the luminous supernovae in the Anderson et al. (2014) sample.}
\label{fig:andcomp}
\end{figure}

In order to check the luminosity fit for consistency, we use the $(g-r)$ color of {\name} at 41 days past explosion (the first date for which we have LCOGT photometry) to calculate a $g$-band bolometric correction of $BC_{g}=-0.63\pm0.09$ based on Lyman, Bersier, \& James (2014). Assuming $E(B-V)=0.05$, we then calculate a bolometric magnitude from the $g$-band magnitude and the bolometric correction and convert this to a bolometric luminosity, obtaining $L_{bol}(t_{exp}+41$~days$)$=$(8.6\pm0.9)\times10^8$~{\lsun}. The bolometric luminosity fit by the PP15 model at this time is $L_{bol}(t_{exp}+41$~days$)$=$(1.4\pm0.2)\times10^9$~{\lsun}, consistent with the Lyman, Bersier, \& James (2014) value to roughly 2 sigma. This difference is not unexpected, as the PP15 model overestimates the $g$-band magnitude in its fit for {\name} (see Figure~5). If we use the value of $E(B-V)=0.28$ that is found by the PP15 model, we obtain $L_{bol}(t_{exp}+41$~days$)$=$(1.3\pm0.9)\times10^9$~{\lsun} using the Lyman, Bersier, \& James (2014) method, indicating that the PP15 fit for $E(B-V)$ is perhaps more consistent with the data.


\begin{figure}[htb]
\centering
\includegraphics[width=0.98\linewidth]{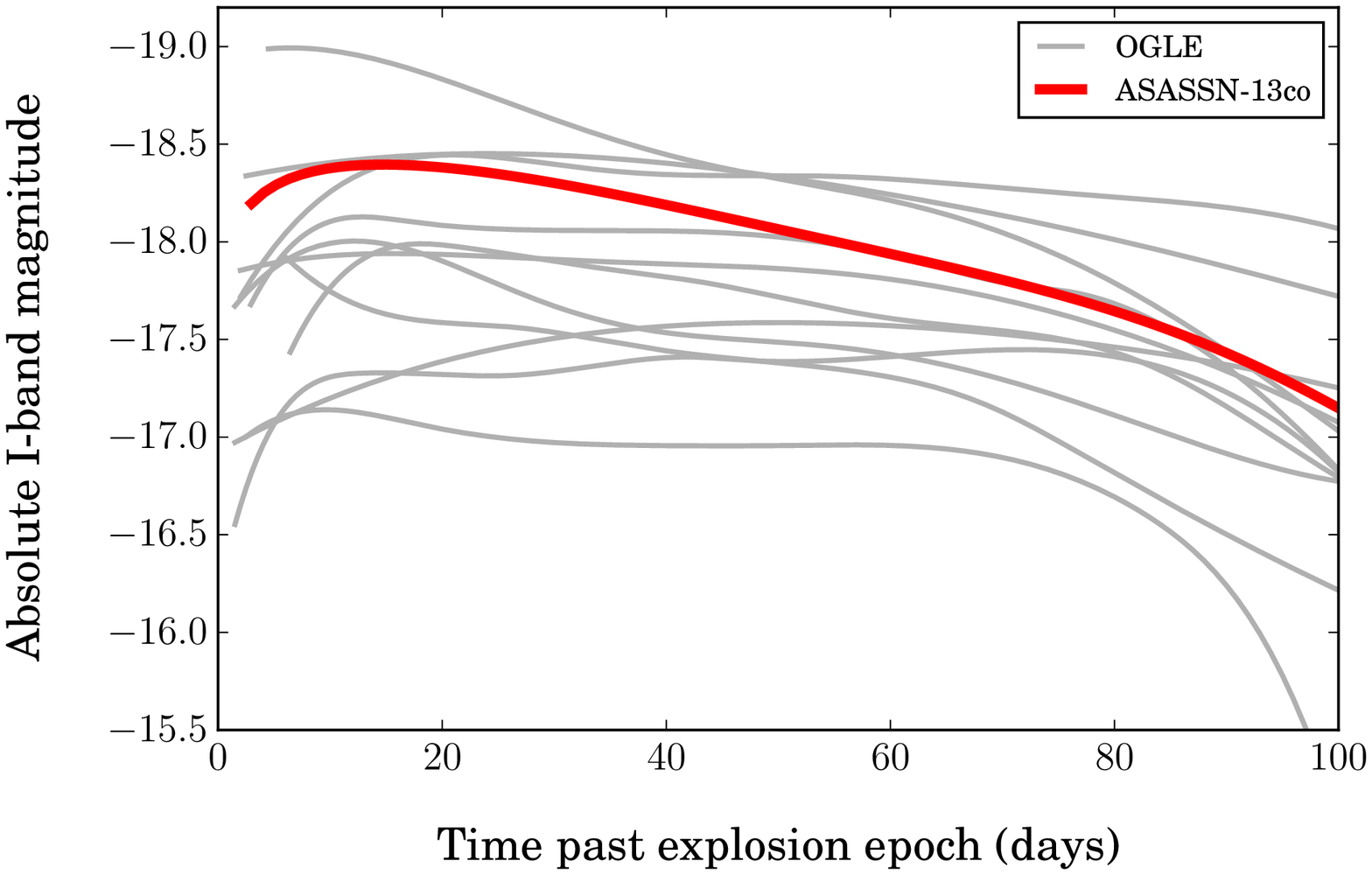}
\FigCap{Comparison of the absolute $I$-band light curve of {\name} (red) to those of the OGLE Type II supernova sample from Poznanski et al. (2015) (grey). The light curve for {\name} is derived from the PP15 model fit and assumes a distance modulus of 34.81. The OGLE light curve fits are spline fits to the $I$-band photometric data. All light curves shown are corrected for Galactic extinction but not for host galaxy extinction. {\name} has one of the most luminous $I$-band light curves of all the supernovae shown, but its decline rate does not seem atypical from the rest of the sample.}
\label{fig:oglecomp}
\end{figure}

Recently there has been some debate about whether Type II supernovae truly fall into distinct ``II-L'' and ``II-P'' groups (e.g., Arcavi et al. 2012; Faran et al. 2014a, b) or whether Type II supernova light curves can have a variety of shapes that do not necessarily conform to one of the two traditional types (e.g., Anderson et al. 2014; Sanders et al. 2014). The distinction may be more apparent in some photometric bands than in others; in particular, \edit{Arcavi et al. (2012)} use $R$-band light curves while Anderson et al. (2014) use $V$-band. In Figure~7 we compare the absolute $V$-band light curve of {\name}, derived from the PP15 model fit, to those of the Type II sample used in Figure 2 of Anderson et al. (2014)\edit{, and in Figure~8 we compare the absolute $I$-band light curve of {\name} to those of the Optical Gravitational Lensing Experiment (OGLE) Type II sample used by Poznanski et al. (2015).} 

\edit{The two Figures illustrate that the II-P or II-L classification can be highly dependent on the photometric band used to observe the supernova. The $V$-band light curve shown in Figure~7 does not seem to be consistent with any of the others in the comparison group: it has a higher absolute magnitude at peak, and although it does appear to decline steadily, like a Type II-L supernova, it appears to do so at an atypically slow rate. While it does not show a distinct ``plateau'', the slow decline appears to be unusual, particularly for a Type II supernova as luminous in the $V$-band as {\name}.}


\begin{figure}[htb]
\centering
\includegraphics[width=0.98\linewidth]{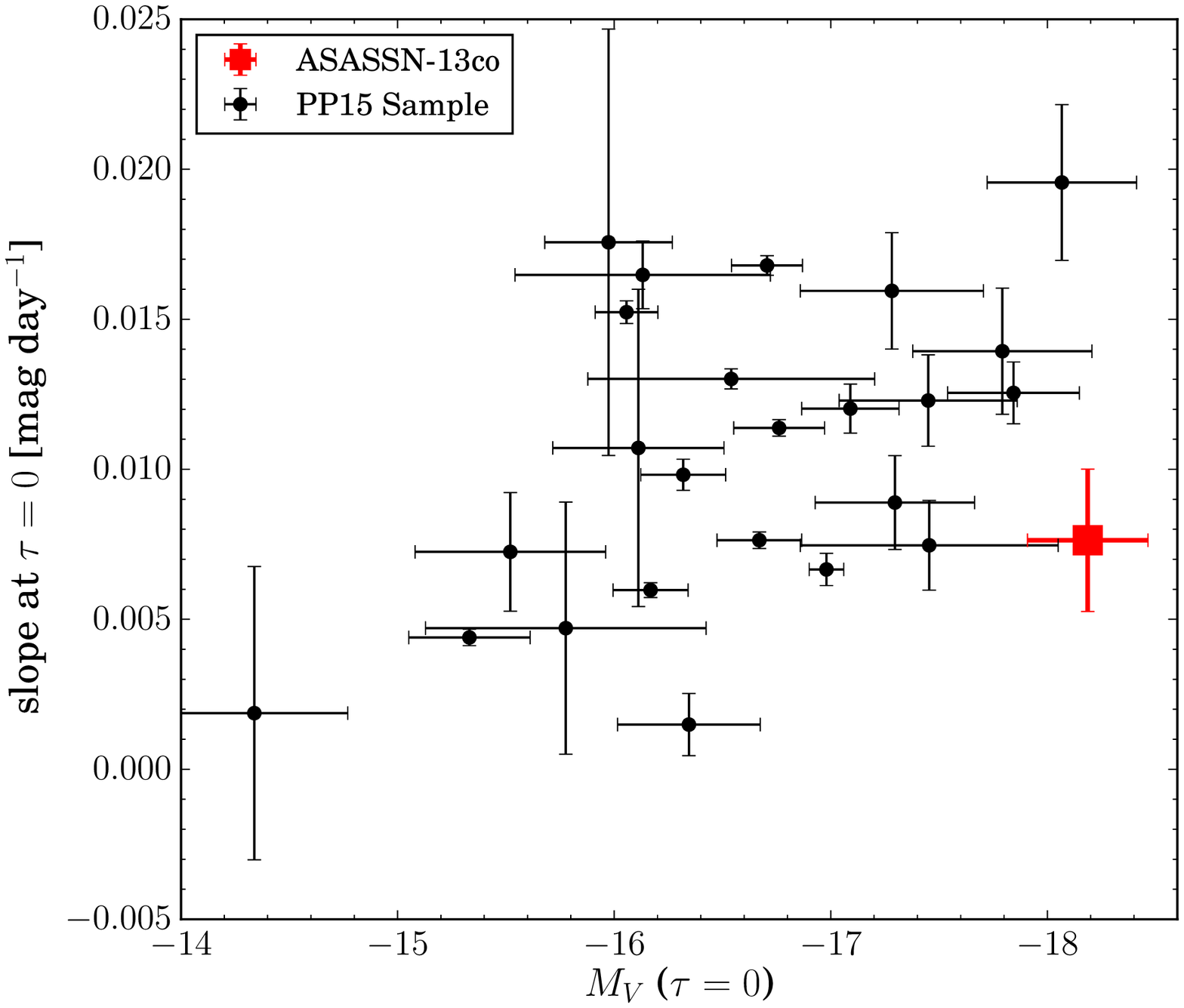}
\FigCap{Correlation of the absolute $V$-band magnitude and the $V$-band light curve slope measured at $\tau=0$ for {\name} (red) and the Type II supernova sample shown in Figure~9 of PP15 (black). {\name} has a shallower slope than other supernovae with similar magnitudes, confirming what can be seen by eye in Figure~7.}
\label{fig:slopemag}
\end{figure}

\edit{Conversely, the $I$-band light curve of {\name} appears to be largely consistent with the OGLE Type II supernova sample shown in Figure~8. To obtain the OGLE light curve fits shown in the Figure, we performed spline fits to the photometric data from Poznanski et al. (2015), while the $I$-band fit for {\name} was again derived from the PP15 model. While {\name} remains one of the more luminous supernovae in the sample, there are other supernovae with similar $I$-band light curve shapes, and the trend of more luminous supernovae declining more rapidly is less apparent in the $I$-band sample than it is in the $V$-band sample. Based on its steady decline rate, {\name} likely would have been classified as an unusually luminous Type II-L supernova based on its $I$-band light curve. That aside, the OGLE sample is composed of only 11 supernovae, and appears to be incomplete at fainter magnitudes. A larger sample spanning a wider magnitude range may well exhibit a more continuous range of decline rates.}

Anderson et al. (2014) found a correlation between the peak absolute $V$-band magnitude and the slope of the $V$-band light curve's decline during the plateau phase, indicating that supernovae with smaller peak magnitudes tend to fade at a faster rate than those with higher peak magnitudes. PP15 were able to reproduce this trend using the absolute $V$-band magnitude and light curve slope at constant effective temperature corresponding to $\tau=0$, corresponding to roughly the middle of the plateau phase, rather than the peak magnitude and ``plateau phase'' slope used by Anderson et al. (2014). In Figure~9 we reproduce Figure 9 from PP15, showing the absolute magnitudes and slopes of the supernovae in their sample, along with {\name}. Figure~9 confirms what can be seen by eye in Figure~7---while {\name} is one of the most luminous supernovae in Figure~7, its \edit{$V$-band} decline slope seems to be more consistent with supernovae that are two or more magnitudes fainter. 

Ultimately, {\name} appears to be photometrically unusual. It is quite luminous for a Type II supernova, with a peak $V$-band absolute magnitude of $\sim-18.1$. Its light curve shape and PP15 model parameters seem to indicate that \edit{while its $I$-band decline rate is fairly typical of a Type II-L supernova, its $V$-band} properties fall somewhere between those of a typical Type II-L and those of a typical Type II-P, perhaps lending credence to the idea that Type II supernovae should not be divided into two distinct groups. However, it also does not conform to the magnitude-slope relations derived in Anderson et al. (2014) or PP15, as its light curve declines at a rate more typical of a supernova that is $\sim2$ magnitudes fainter. While it is spectroscopically very similar to many Type II-P supernovae, \name's photometric properties are inconsistent with that designation.


\section{Discussion}
\label{sec:disc}

Our observations indicate that {\name} was an unusual event. Its spectra match well to normal Type II-P supernovae, but its light curve shape is a better match to that of a Type II-L supernova, and its peak magnitude ($M_V\sim-18.1$) is quite high. Light curve fits with the PP15 model corroborate this picture, showing that the bolometric luminosity was atypically high and that its light curve declined at a slow but steady rate. Because it does not fit into either of the traditional designations, we conclude that {\name} provides additional evidence that Type II supernovae exhibit a wide range of physical and observed properties that cannot be quantified by the simple distinction between II-P and II-L. \edit{However, this distinction may also depend on the photometric bands used for follow-up observation. More observational evidence in multiple photometric bands will be required to definitively determine whether the II-P/II-L distinction is real, or whether Type II supernovae simply exhibit a broad but continuous range of photometric properties.}

{\name} has proven to be an example of the usefulness of both the PP15 model and the ASAS-SN project. While the PP15 model was designed specifically to model the light curves of Type II-P supernovae, it is able to provide a reasonable fit to our data despite {\name} exhibiting a light curve that declines at an unusually slow rate. That we are able to extract a bolometric luminosity curve and other properties from these fits that do not depend on theoretical models provides an example of how the PP15 model will be a powerful tool for investigating the diverse properties of Type II supernovae in the future. It also allows us to compare objects with different photometric datasets, as we have done in our comparison with the Anderson et al. (2014) and Poznanski et al. (2015) samples: {\name} has no Johnson $V$- or $I$-band data, but the PP15 model allows us to ``predict'' that flux, making a comparison possible.

{\name} represents a similar success for the ASAS-SN project, as it is an unusual event and provides an example of the quality of study that can be done with ASAS-SN discoveries. As ASAS-SN is comprised only of small telescopes, its discoveries can be observed in great detail using relatively small 1- and 2-m class telescopes, and with 296 supernova discoveries (including 62 Type II supernovae) at time of writing, it will undoubtedly be an extremely useful project for investigating the diversity of these events in the future.

\Acknow{The authors thank PI Neil Gehrels and the {\swift} ToO team for promptly approving and executing our observations and J. S. Brown for assistance with host property fitting. We thank LCOGT and its staff for their continued support of ASAS-SN.

ASAS-SN is supported by NSF grant AST-1515927. Development of ASAS-SN has been supported by NSF grant AST-0908816, the Center for Cosmology and AstroParticle Physics at the Ohio State University, the Mt. Cuba Astronomical Foundation, and by George Skestos.

TW-SH is supported by the DOE Computational Science Graduate Fellowship, grant number DE-FG02-97ER25308. Support for JLP is in part provided by FONDECYT through the grant 1151445 and by the Ministry of Economy, Development, and Tourism's Millennium Science Initiative through grant IC120009, awarded to The Millennium Institute of Astrophysics, MAS. OP is supported by NASA through Hubble Fellowship grant HST-HF-51327.01-A awarded by the Space Telescope Science Institute, which is operated by the Association of Universities for Research in Astronomy, Inc., for NASA, under contract NAS 5-26555. BS is supported by NASA through Hubble Fellowship grant HST-HF-51348.001 awarded by the Space Telescope Science Institute, which is operated by the Association of Universities for Research in Astronomy, Inc., for NASA, under contract NAS 5-26555. JFB is supported by NSF grant PHY-1404311.

This research has made use of the XRT Data Analysis Software (XRTDAS) developed under the responsibility of the ASI Science Data Center (ASDC), Italy. At Penn State the NASA {\swift} program is support through contract NAS5-00136.

This research was made possible through the use of the AAVSO Photometric All-Sky Survey (APASS), funded by the Robert Martin Ayers Sciences Fund.

This research has made us of data provided by Astrometry.net (Barron et al. 2008).

This research has made use of the NASA/IPAC Extragalactic Database (NED), which is operated by the Jet Propulsion Laboratory, California Institute of Technology, under contract with the National Aeronautics and Space Administration.}

\appendix

\section*{Appendix. Follow-Up Photometry and Spectroscopic Velocities}

All host-subtracted follow-up photometry and photospheric velocities are presented in Table~2 and Table~3 below, respectively. Photometry is presented in the natural system for each filter: $ugriz$ magnitudes are in the AB system, while {\swift} filter magnitudes are in the Vega system.

\MakeTable{cccc|cccc}{12.5cm}{Host-subtracted photometric data of {\name}}
{\hline
MJD & Magnitude &  Filter & Telescope & MJD & Magnitude &  Filter & Telescope\\
\hline
56569.738 & 17.212 0.045 & $i$ & LCOGT & 56543.407 & 16.434 0.104 & $U$ & {\swift}\\ 
56582.811 & 17.246 0.040 & $i$ & LCOGT & 56547.478 & 16.972 0.156 & $U$ & {\swift}\\ 
56582.821 & 17.245 0.043 & $i$ & LCOGT & 56550.289 & 17.389 0.243 & $U$ & {\swift}\\ 
56600.058 & 17.751 0.058 & $i$ & LCOGT & 56555.614 & $>$17.478 & $U$ & {\swift}\\ 
56617.036 & 17.936 0.069 & $i$ & LCOGT & 56559.495 & $>$17.596 & $U$ & {\swift}\\ 
56624.048 & 18.231 0.097 & $i$ & LCOGT & 56563.824 & $>$18.504 & $U$ & {\swift}\\ 
56640.093 & 19.076 0.232 & $i$ & LCOGT & 56568.098 & $>$17.844 & $U$ & {\swift}\\ 
56641.115 & 18.832 0.179 & $i$ & LCOGT & 56572.231 & $>$18.270 & $U$ & {\swift}\\ 
56569.735 & 17.135 0.023 & $r$ & LCOGT & 56576.436 & $>$19.035 & $U$ & {\swift}\\ 
56582.808 & 17.451 0.026 & $r$ & LCOGT & 56581.718 & $>$18.552 & $U$ & {\swift}\\ 
56582.818 & 17.429 0.025 & $r$ & LCOGT & 56584.444 & $>$19.605 & $U$ & {\swift}\\ 
56600.055 & 17.762 0.027 & $r$ & LCOGT & 56538.942 & 16.513 0.101 & $W1$ & {\swift}\\ 
56617.033 & 17.899 0.031 & $r$ & LCOGT & 56541.669 & 16.935 0.119 & $W1$ & {\swift}\\ 
56624.045 & 18.091 0.038 & $r$ & LCOGT & 56543.404 & 17.362 0.159 & $W1$ & {\swift}\\ 
56640.090 & 18.871 0.102 & $r$ & LCOGT & 56547.476 & 18.133 0.295 & $W1$ & {\swift}\\ 
56641.112 & 19.137 0.124 & $r$ & LCOGT & 56550.287 & $>$17.868 & $W1$ & {\swift}\\ 
56538.950 & 16.916 0.220 & $V$ & {\swift} & 56555.611 & $>$18.681 & $W1$ & {\swift}\\ 
56541.680 & 16.660 0.174 & $V$ & {\swift} & 56559.492 & $>$18.652 & $W1$ & {\swift}\\ 
56543.414 & 16.834 0.194 & $V$ & {\swift} & 56563.821 & $>$18.393 & $W1$ & {\swift}\\ 
56550.293 & 17.372 0.617 & $V$ & {\swift} & 56568.096 & $>$18.488 & $W1$ & {\swift}\\ 
56555.621 & 16.916 0.209 & $V$ & {\swift} & 56572.229 & $>$18.388 & $W1$ & {\swift}\\ 
56559.501 & 17.004 0.239 & $V$ & {\swift} & 56576.434 & $>$18.900 & $W1$ & {\swift}\\ 
56563.831 & 17.227 0.278 & $V$ & {\swift} & 56581.716 & $>$18.501 & $W1$ & {\swift}\\ 
56568.102 & 16.818 0.212 & $V$ & {\swift} & 56584.442 & $>$18.133 & $W1$ & {\swift}\\ 
56572.236 & 17.060 0.265 & $V$ & {\swift} & 56538.951 & 16.915 0.157 & $M2$ & {\swift}\\ 
56576.440 & 17.509 0.381 & $V$ & {\swift} & 56541.681 & 17.604 0.129 & $M2$ & {\swift}\\ 
56581.723 & 17.183 0.297 & $V$ & {\swift} & 56543.415 & 17.902 0.165 & $M2$ & {\swift}\\ 
56584.448 & 17.480 0.371 & $V$ & {\swift} & 56550.756 & 17.961 0.563 & $M2$ & {\swift}\\ 
56569.733 & 18.037 0.065 & $g$ & LCOGT & 56555.622 & $>$18.474 & $M2$ & {\swift}\\ 
56582.816 & 18.362 0.051 & $g$ & LCOGT & 56559.502 & $>$18.762 & $M2$ & {\swift}\\ 
56600.053 & 18.991 0.070 & $g$ & LCOGT & 56563.832 & $>$18.434 & $M2$ & {\swift}\\ 
56624.042 & 19.961 0.182 & $g$ & LCOGT & 56568.103 & $>$18.636 & $M2$ & {\swift}\\ 
56641.110 & 20.546 0.388 & $g$ & LCOGT & 56572.237 & $>$18.430 & $M2$ & {\swift}\\ 
56538.945 & 16.944 0.104 & $B$ & {\swift} & 56576.441 & $>$18.400 & $M2$ & {\swift}\\ 
56541.674 & 17.129 0.114 & $B$ & {\swift} & 56581.723 & $>$18.773 & $M2$ & {\swift}\\ 
56543.408 & 17.129 0.114 & $B$ & {\swift} & 56584.449 & $>$18.676 & $M2$ & {\swift}\\ 
56547.479 & 17.141 0.134 & $B$ & {\swift} & 56538.946 & 17.407 0.122 & $W2$ & {\swift}\\ 
56550.289 & 17.338 0.149 & $B$ & {\swift} & 56541.675 & 17.719 0.151 & $W2$ & {\swift}\\ 
56555.615 & 17.444 0.142 & $B$ & {\swift} & 56543.409 & 18.159 0.212 & $W2$ & {\swift}\\ 
56559.496 & 17.755 0.187 & $B$ & {\swift} & 56550.290 & $>$18.472 & $W2$ & {\swift}\\ 
56563.825 & 18.017 0.238 & $B$ & {\swift} & 56555.616 & $>$19.275 & $W2$ & {\swift}\\ 
56568.099 & 18.405 0.360 & $B$ & {\swift} & 56559.497 & $>$19.188 & $W2$ & {\swift}\\ 
56572.232 & 18.607 0.424 & $B$ & {\swift} & 56563.826 & $>$18.675 & $W2$ & {\swift}\\ 
56576.437 & 18.472 0.389 & $B$ & {\swift} & 56568.099 & $>$18.709 & $W2$ & {\swift}\\ 
56581.719 & $>$18.168 & $B$ & {\swift} & 56572.233 & $>$18.735 & $W2$ & {\swift}\\ 
56584.444 & $>$17.839 & $B$ & {\swift} & 56576.437 & $>$18.833 & $W2$ & {\swift}\\ 
56538.944 & 16.067 0.083 & $U$ & {\swift} & 56581.720 & $>$18.787 & $W2$ & {\swift}\\ 
56541.673 & 16.325 0.097 & $U$ & {\swift} & 56584.445 & $>$18.534 & $W2$ & {\swift}\\
\hline
\multicolumn{8}{p{12.5cm}}{Magnitudes and uncertainties are presented in the natural system for each filter: $ugriz$ magnitudes are presented in the AB system, {\swift} filter magnitudes are presented in the Vega system. 3-sigma upper limits are given for epochs with no detection.}
\label{table:phot}
}

\MakeTable{ccccccc}{12.5cm}{Photospheric velocities of {\name}}
{\hline
 & H$\beta$ & \ion{Fe}{2} & \ion{He}{1} & \ion{Si}{2} & H$\alpha$ & \ion{O}{1} \\
 MJD & (4861~\AA) & (5169~\AA) & (5875~\AA) & (6355~\AA) & (6563~\AA) & (7773~\AA) \\
\hline
56536.10 & $-11478$ & ---  & $-11220$ & $-9562$ & $-11603$ & ---\\
56537.14 & $-11576$ & ---  &   ---  & --- & --- & ---\\ 
56538.10 & $-11040$ & ---  & $-11230$ & $-9468$ & $-11863$ & ---\\ 
56539.18 & $-10831$ & ---  & $-11399$ & $-9765$ & $-11297$ & ---\\ 
56540.14 & $-11589$ & $-9356$ & $-11159$ & $-8930$ & $-11219$ & ---\\
56546.26 & $-10269$ & $-8885$ & $-10195$ & $-9142$ & $-11114$ & ---\\
56552.29 & $-9647$ & $-8615$ & $-9794$ & $-8270$ & $-11022$ & ---\\
56591.22 & $-7273$ & $-5455$ & $-6528$ & --- & $-8544$ & $-6213$\\
56611.08 & --- & $-4599$ & $-6128$ & --- & $-8153$ & ---\\
\hline
\multicolumn{7}{p{10cm}}{Column headers indicate the lines and wavelengths used to measure the velocities. All velocities are listed in km~s$^{-1}$ and no value is given for epochs where a measurement was not possible.}
\label{table:phot}
}


\begin{references}
\refitem{Ahn, C.~P., et al.}{2012}{ApJS}{203}{21}
\refitem{Alard, C.}{2000}{AAPS}{144}{363}
\refitem{Anderson, J.~P., et al.}{2014}{ApJ}{786}{67}
\refitem{Arcavi, I., et al.}{2012}{ApJL}{756}{L30}
\refitem{Barron, J. T., et al.}{2008}{AJ}{135}{414}
\refitem{Blondin, S., Tonry, J.~L.}{2007}{ApJ}{666}{1024}
\refitem{Breeveld, A.~A., et al.}{2010}{MNRAS}{406}{1687}
\refitem{Brown, T.~M., et al.}{2013}{PASP}{125}{1031}
\refitem{Bruzual, G., Charlot, S.}{2003}{MNRAS}{344}{1000}
\refitem{Burrows, D.~N., et al.}{2005}{SSR}{120}{165}
\refitem{Cardelli, J.~A., Clayton, G.~C., Mathis, J.~S.}{1988}{ApJL}{329}{L33}
\refitem{Faran, T., et al.}{2014a}{MNRAS}{442}{844}
\refitem{Faran, T., et al.}{2014b}{MNRAS}{445}{554}
\refitem{Filippenko, A.~V.}{1997}{ARA\&A}{35}{309}
\refitem{Hill, J.~E.}{2004}{SPIE Conference Series, X-Ray and Gamma-Ray Instrumentation for Astronomy XIII}{5165}{217}
\refitem{Holoien, T.~W.-S., et al.}{2013}{The Astronomer's Telegram}{5346}{1}
\refitem{Holoien, T.~W.-S., et al.}{2014a}{MNRAS}{445}{3263}
\refitem{Holoien, T.~W.-S., et al.}{2014b}{ApJL}{785}{L35}
\refitem{Holoien, T.~W.-S., et al.}{2016a}{MNRAS}{455}{2918}
\refitem{Holoien, T.~W.-S., et al.}{2016b}{~}{~}{arXiv:1602.01088}
\refitem{Kalberla, P.~M.~W., et al.}{2005}{AAP}{440}{775}
\refitem{Kriek, M., et al.}{2009}{ApJ}{700}{221}
\refitem{Leonard, D.~C., et al.}{2002}{AJ}{124}{2490}
\refitem{Lyman, J.~D, Bersier, D., James, P. A.}{2014}{MNRAS}{437}{3848}
\refitem{Martini, P., et al.}{2011}{PASP}{123}{187}
\refitem{Morrell, N., Prieto, J.~L.}{2013}{The Astronomer's Telegram}{5353}{1}
\refitem{Quimby,R.~M., et al.}{2007}{ApJ}{666}{1093}
\refitem{Sanders, N.~E., et al.}{2015}{ApJ}{799}{208}
\refitem{Skrutskie, M.~F., et al.}{2006}{AJ}{131}{1163}
\refitem{Pejcha, O., Prieto, J.~L.}{2015a}{ApJ}{799}{215}
\refitem{Pejcha, O., Prieto, J.~L.}{2015b}{ApJ}{806}{225}
\refitem{Poole, T.~S., et al.}{2008}{MNRAS}{383}{627}
\refitem{Poznanski, D., et al.}{2015}{MNRAS}{449}{1753}
\refitem{Roming, P.~W.~A., et al.}{2005}{SSR}{120}{95}
\refitem{Sanders, N.~E., et al.}{2015}{ApJ}{799}{208}
\refitem{Schmidt., S.~J. et al.}{2014}{ApJL}{781}{L24}
\refitem{Shappee, B.~J., et al.}{2014}{ApJ}{788}{48}
\refitem{Springob, C.~M. et al.}{2005}{ApJS}{160}{149}
\refitem{Valenti, S., et al.}{2014}{MNRAS}{438}{L101}
\refitem{Voges, W., et al.}{1999}{AAP}{349}{389}
\refitem{Yaron, O., Gal-Yam, A.}{2012}{PASP}{124}{668}
\end{references}
\end{document}